\documentclass[sigconf,natbib=true]{acmart}

\AtBeginDocument{%
  }

\copyrightyear{2025}
\acmYear{2025}
\setcopyright{cc}
\setcctype{by}
\acmConference[SIGIR '25]{Proceedings of the 48th International ACM SIGIR Conference on Research and Development in Information Retrieval}{July 13--18, 2025}{Padua, Italy}
\acmBooktitle{Proceedings of the 48th International ACM SIGIR Conference on Research and Development in Information Retrieval (SIGIR '25), July 13--18, 2025, Padua, Italy}\acmDOI{10.1145/3726302.3730348}
\acmISBN{979-8-4007-1592-1/2025/07}

\usepackage{booktabs}
\usepackage{multirow}
\usepackage{xcolor}

\usepackage{enumitem}
\setlist[itemize]{leftmargin=*}

\makeatletter
\newcommand{\issue}[2]{%
  \vspace{\baselineskip}
  \par\noindent
  {\bfseries \itshape Issue \##1: #2} 
}
\makeatother

\makeatletter
\newcommand{\boldheading}[1]{%
  \vspace{\baselineskip}
  \par\noindent
  {\bfseries #1} 
}
\makeatother

\begin{document}
\fancyhead{}

\title{Rankers, Judges, and Assistants: Towards Understanding the Interplay of LLMs in Information Retrieval Evaluation}

\author{Krisztian Balog}
\affiliation{%
  \institution{Google DeepMind}
  \city{Stavanger}
  \country{Norway}
}
\email{krisztianb@google.com}

\author{Donald Metzler}
\affiliation{%
  \institution{Google DeepMind}
  \city{Mountain View}
  \country{USA}
}
\email{metzler@google.com}

\author{Zhen Qin}
\affiliation{%
  \institution{Google DeepMind}
  \city{Mountain View}
  \country{USA}
}
\email{zhenqin@google.com}

\begin{abstract}
Large language models (LLMs) are increasingly integral to information retrieval (IR), powering ranking, evaluation, and AI-assisted content creation. This widespread adoption necessitates a critical examination of potential biases arising from the interplay between these LLM-based components. 
This paper synthesizes existing research and presents novel experiment designs that explore how LLM-based rankers and assistants influence LLM-based judges.
We provide the first empirical evidence of LLM judges exhibiting significant bias towards LLM-based rankers. Furthermore, we observe limitations in LLM judges' ability to discern subtle system performance differences. Contrary to some previous findings, our preliminary study does not find evidence of bias against AI-generated content.
These results highlight the need for a more holistic view of the LLM-driven information ecosystem. To this end, we offer initial guidelines and a research agenda to ensure the reliable use of LLMs in IR evaluation.
\end{abstract}



\maketitle

\section{Introduction}

Due to their remarkable capabilities, large language models (LLMs) are fundamentally reshaping the field of information retrieval (IR), becoming integral to core ranking algorithms and the automation of evaluation processes. Beyond their role in core IR processes, LLMs are also powering AI assistants that are rapidly changing how users generate content, from writing emails and articles to creating code and translating content between languages. As the reliance on LLMs is expected to deepen given their potential, it is increasingly crucial to maintain a balanced perspective by assessing and acknowledging the potential risks alongside the undeniable benefits. Could this heavy reliance on LLMs across content creation, retrieval, ranking, evaluation, etc., inadvertently introduce or amplify biases within these systems?

Recent research has begun to explore some of these emerging issues. For example, studies have shown that LLMs can exhibit biases in their output, favoring LLM-generated content over human-generated ones~\citep{Dai:2024:KDD}, and perpetuating biases present in their training data~\citep{Gallegos:2024:CL,Liang:2021:ICML,Navigli:2023:JCIQ}.  Furthermore, LLM-based rating systems have been found to be susceptible to manipulation~\citep{Alaofi:2024:SIGIRAP}, may not accurately reflect human preferences~\citep{Liu:2024:COLM}, and demonstrate self-inconsistency~\citep{Stureborg:2024:arXiv}. Additionally, the phenomenon of ``model collapse'' has also been observed, where LLMs trained on synthetic data generated by other LLMs can lead to a degradation of quality and diversity in generated content~\citep{Shumailov:2024:Nature}.

Within the IR research community, the use of LLMs for assessment is a subject of ongoing debate~\citep{Faggioli:2023:ICTIR}, with opinions ranging from complete rejection of LLMs for relevance assessment~\citep{Soboroff:2025:IRR} to the assertion that they can fully replace human judgments~\citep{Upadhyay:2024b:arXiv}.
Investigations have thus far focused on the agreement of LLM-generated ratings with human assessments~\citep{Faggioli:2023:ICTIR,Upadhyay:2024b:arXiv,Thomas:2024:SIGIR} and the potential for LLMs to introduce biases in search results~\citep{Dai:2024:KDD}. However, a comprehensive analysis of the implications of LLMs across the entire information ecosystem, from content creation with AI assistance to LLM-based reranking and LLM-based judges for evaluation, remains a critical gap in the current literature.

This paper aims to advance our understanding of these issues by synthesizing prior research and, crucially, providing novel empirical evidence. 
We specifically focus on the novel challenge of understanding the effect LLM-based rankers and AI-powered content creation have on an LLM-based judge's ability to accurately assess relevance. 
Prior work has separately noted the \emph{potential} interaction between LLM-based rankers and judges~\citep{Rahmani:2024:arXiv,Faggioli:2023:ICTIR,Thomas:2024:SIGIR,MacAvaney:2023:SIGIR} (an interaction that has yet to be empirically investigated), while other initial work has explored the relationship between AI-powered content creation and rankers~\citep{Dai:2024:KDD}.
However, we argue that the complex interplay between each of these roles must be considered holistically to fully understand the potential implications of widespread adoption of LLM-based judges.
We present initial results demonstrating the importance of this interconnected perspective, showcasing how the use of LLMs across the information lifecycle can influence the accuracy and potential biases of LLM judges.

We start by considering the case of LLMs being used as both rankers and judges and present the first empirical demonstration of a significant bias of LLM judges towards LLM-based rankers.
Novel to our approach is the examination of LLM judge performance via the use of oracle rankers, allowing for a controlled assessment of LLM judge behavior and discriminative ability. 
Using the TREC 2019 and 2020 Deep Learning track datasets, we conduct experiments that also compare different-sized LLM judges within the same model family.  
Our results reveal several key findings: (1) LLM judges are more lenient in their relevance assessments than human judges, confirming previous observations~\citep{Upadhyay:2024b:arXiv}; (2) LLM judges exhibit a significant bias towards LLM-based rankers, a phenomenon previously only hypothesized; and (3) LLM judges demonstrate limited ability to discern subtle, yet statistically significant, performance differences between systems.
Additionally, we conduct a preliminary study into whether LLM judges demonstrate biases when they encounter AI-generated content. Contrary to some previously published findings~\citep{Liu:2023:EMNLP,Panickssery:2024:NeurIPS,Liu:2024:ACL}, our experiments do \emph{not} provide evidence of this bias, suggesting that deeper, more rigorous empirical evaluations are required to better understand this phenomenon. 

What emerges from these targeted studies is a better picture of how different interactions between LLM-based components give rise to different behaviors within LLM-based judges. Taken together, our findings yield one of the most holistic views of this problem space, provide unique insights into best practices for leveraging LLMs as judges, and motivate a rich set of future research questions that will need to be answered to understand the complexities of these interactions even better.

In summary, the primary contributions of this paper are:
(1) a review of how LLMs are currently used in IR, bringing attention to the interconnected roles they play, and synthesizing the current understanding of their interactions;
(2) experiments that highlight how interactions between LLMs might result in inaccurate or biased assessments of retrieval effectiveness;
(3) a preliminary set of guidelines for using LLMs in IR evaluation; and
(4) a research agenda aimed at sparking further discussion and research along this emerging direction.

\section{Background}

\begin{figure}[t]
    \centering
    \includegraphics[width=0.45\textwidth]{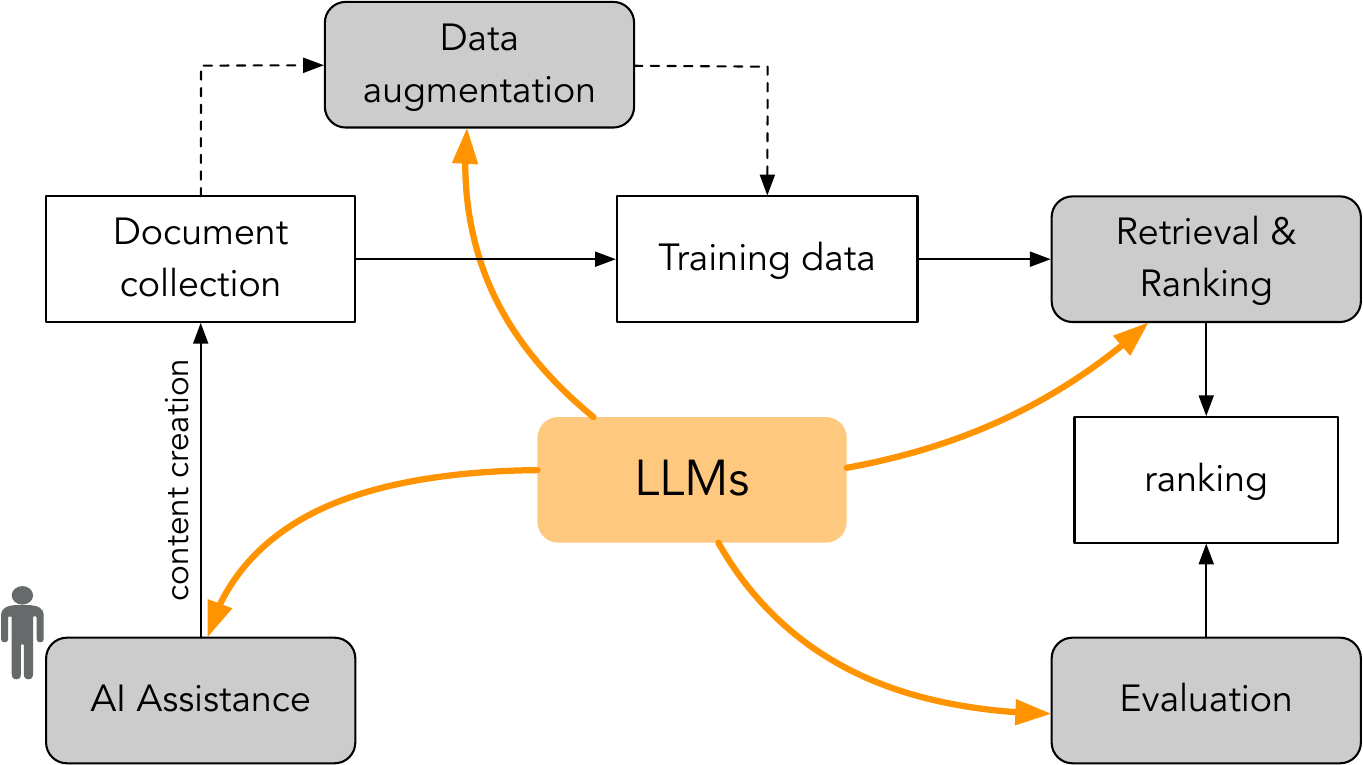}
    \caption{LLM usage in modern information access systems.}
    \label{fig:overview}
    \vspace*{-0.5\baselineskip}
\end{figure}

This section overviews the main uses of LLMs in information access, illustrated in Fig.~\ref{fig:overview}, providing context for subsequent analysis of the interplay between some of these uses.

\paragraph{\textbf{LLMs as Rankers}}

In modern large-scale IR systems, a multi-stage retrieve-then-rerank pipeline has become a prominent approach, wherein an initial retrieval stage, often based on lexical matching or embedding-based methods, is followed by one or multiple reranking stages, utilizing more sophisticated models to refine the results. This reranking stage frequently employs LLMs, either fine-tuned for the task of ranking~\citep{Nogueira:2020:EMNLP,Zhuang:2023:SIGIR,Nogueira:2019:arXiv,Pradeep:2021:arXiv} or via prompting in a pointwise~\citep{Liang:2023:TMLR,Drozdov:2023:EMNLP,Sachan:2022:EMNLP}, pairwise~\citep{Qin:2024:NAACL}, or listwise~\citep{Sun:2023:EMNLP,Ma:2023:arXiv} fashion.
\citet{Dai:2024:KDD} present results suggesting an inherent bias in neural retrieval models toward LLM-generated texts. This \emph{source bias} may stem from shared Transformer-based architectures and pretraining approaches, and can lead to ``semantic shortcuts'' during matching.
Neural IR models are also shown to be vulnerable to adversarial attacks, such as keyword stuffing and content injection~\citep{Parry:2024:ECIR,Tamber:2025:arXiv}.

\paragraph{\textbf{LLMs as Judges}}

Early LLMs, such as BERT, have been utilized for measuring the distributional similarity of texts~\citep{Zhang:2020:ICLR,Zhao:2019:EMNLP-IJCNLP} and for evaluating specific tasks via fine-tuning, including machine translation~\citep{Zhu:2020:ICLR}, text summarization~\citep{Liu:2019:EMNLP}, and question answering~\citep{McCarley:2019:arXiv}.
The arrival of generative LLMs, such as ChatGPT, have enabled various data labeling and annotation tasks~\citep{Gilardi:2023:PNAS}. The use of LLMs as surrogates for humans for evaluation, often referred to as ``LLM-as-a-Judge''~\citep{Zheng:2023:NeurIPS}, now extends across virtually all natural language processing tasks, including text summarization and dialog response generation~\citep{Gu:2024:arXiv}. However, recent research increasingly demonstrates their limitations, such as favoring longer responses (\emph{length bias})~\citep{Dubois:2024:COLM,Wang:2023:NeurIPS} or content generated by similar models (\emph{self bias})~\citep{Liu:2024:ACL,Xu:2024:ACL}.
Our interest is specifically in the use of LLMs for relevance assessments in IR.
\citet{MacAvaney:2023:SIGIR} is among the first to employ LLMs for automatic relevance labeling. They specifically focus on a setting where a single known relevant document per query is available for evaluation and explore several one-shot approaches. 
\citet{Faggioli:2023:ICTIR} present a spectrum of human-machine collaboration for producing relevance assessments, from AI assistance to fully automated judgments. They conduct a preliminary assessment of LLMs' capabilities of relevance judgments on two TREC collections and report a fair agreement between human assessors and LLMs. 
\citet{Thomas:2024:SIGIR} experiment with various prompt templates to improve quality and observe better agreement with the official TREC labels than \citet{Faggioli:2023:ICTIR}. These improvements are attributed to both prompt design and the use of a more capable LLM. 
\citet{Thomas:2024:SIGIR} further share experiences on using LLMs for relevance assessment at Microsoft Bing, where LLMs have reportedly been used, in conjunction with expert human labelers, since late 2022. 
\citet{Upadhyay:2024:arXiv} reproduce results from \citet{Thomas:2024:SIGIR}, verifying their claims, and create an open-source implementation (UMBRELA).
Most recently, LLMs are leveraged in the TREC 2024 Retrieval Augmented Generation (RAG) track for automatic relevance assessment~\citep{Upadhyay:2024b:arXiv}.
Relative system rankings are found to correlate with those obtained using human judgments, even if human assessors apply stricter relevance criteria than LLMs~\citep{Upadhyay:2024b:arXiv}.
The authors also experiment with various LLM-assisted labeling processes, such as using UMBRELA to pre-filter the pools or to suggest relevance  labels that human judges can then post-edit, but find that those solutions ``do not appear to have obvious tangible benefits over fully automatic processes''~\citep{Upadhyay:2024b:arXiv}.
\citet{Clarke:2024:arXiv} raise concerns about the claims made by \citet{Upadhyay:2024b:arXiv} and highlight how LLM-based judgments fail to demonstrate strong alignment with manual judgments for top-performing systems. They further present evidence that when evaluation is performed through a publicly known automatic process, such as UMBRELA, it can be subject to manipulation. 
\citet{Chen:2024:SIGIRAP} show that when performing relevance assessments in batches, the relevance levels of earlier documents in a batch influences the relevance judgments of subsequent documents, and that some LLMs are more affected by this so-called \emph{threshold priming effect} than others.
\citet{Alaofi:2024:SIGIRAP} compare various open-source and proprietary LLMs in labeling passages for relevance. 
They demonstrate that most LLMs exhibit some degree of susceptibility to judging non-relevant documents as relevant if query words are inserted at random positions, simulating a keyword stuffing SEO strategy.
\citet{Rahmani:2025:arXiv} present a large-scale synthetic passage ranking collection, SnyDL, by extending the TREC 2019-2023 Deep Learning collections via LLM-generated labels, and observe a high agreement on system ordering.

\paragraph{\textbf{LLMs as Assistants}}

There is a wide array of AI tools available to aid people with content creation.  Focusing only on textual content, the spectrum ranges from basic grammar and spell checkers to advanced tools that generate full articles. 
Studies indicate that by late 2024, LLM assistance is detectable in a significant portion of various text domains, with estimates reaching up to 18\% of financial consumer complaints and 24\% in corporate press releases~\citep{Liang:2025:arXiv}.
The use of powerful LLMs can lead to situations where it is unclear whether the content is primarily human-created with AI assistance or the other way around.

\paragraph{\textbf{LLMs for Data Augmentation}}

While not considered for this role in the current paper, LLMs are also used for data augmentation. For example, \citet{Dai:2023:ICLR} use few-shot prompting to generate synthetic queries, while  \citet{Bonifacio:2022:SIGIR} consider query generation in a full unsupervised setting. \citet{Soudani:2024:arXiv} present a survey on synthetic dialogue data generation in open-domain, task-oriented, and information seeking dialogue systems. 
The use of LLM-generated data brings forth new challenges in bias and unfairness, potentially affecting the reliability of IR systems~\citep{Dai:2024b:KDD}.

\section{Critical Issues with LLMs as Judges}
\label{sec:issues}

While LLMs offer promising capabilities for automated evaluation in IR, a growing body of research highlights potential limitations and raises critical concerns about their widespread adoption as judges. This section synthesizes findings from prior work, identifying key challenges that warrant further investigation. We categorize these challenges into two broad areas: the quality of LLM judgments (Section~\ref{sec:issues:quality}) and the vulnerability of LLM judges to bias and manipulation (Section~\ref{sec:issues:bias}). Within these areas, we discuss specific issues related to validity, discriminative power, reliability, reproducibility, susceptibility to manipulation, and systemic biases. These issues, if unaddressed, could undermine the integrity of IR evaluation and potentially lead to misleading conclusions about system performance. 
This section discusses these critical issues, while Section~\ref{sec:experiments} presents initial experiments designed to provide empirically-driven insight into each of the issues and Section~\ref{sec:challenges} touches upon the fundamental issue of whether, and how, LLM judges should be used in practice.

\subsection{Quality of Judgments}
\label{sec:issues:quality}

The fundamental question underlying the use of LLMs as judges is whether their judgments accurately reflect ``true'' relevance and effectively differentiate between systems of varying quality. We break down this question of quality along two sub-dimensions: \emph{validity and discriminative power} and \emph{reliability and reproducibility}.

\subsubsection{Validity and Discriminative Power}
\label{sec:issues:validity}

For LLMs to serve as effective judges, their assessments must align with human judgments of relevance. 
Existing research measures this in two ways: (1) agreement on individual document-query relevance labels and (2) agreement on the relative ranking of a set of systems.

\begin{itemize}
    \item \emph{Agreement on Individual Relevance Judgments:}
Several studies demonstrated that it is possible to use LLMs for relevance assessment and obtain performance comparable to TREC judges~\citep{Faggioli:2023:ICTIR,Upadhyay:2024b:arXiv} and notably better than crowd judges~\citep{Thomas:2024:SIGIR}. 
At the same time, it has also been observed that LLMs are more lenient when labeling a document relevant~\citep{Upadhyay:2024b:arXiv,Alaofi:2024:SIGIRAP}, which leads to inflated evaluation scores. This leniency can lead to inflated evaluation scores, potentially masking subtle differences between systems.

    \item \emph{Agreement on System Rankings:}
A common approach to meta-evaluating LLM judges is to compare the relative ranking of retrieval systems based on LLM assessments with the ranking based on human-generated relevance judgments. This typically involves calculating the correlation between the two rankings, often using systems submitted to TREC tracks \citep{MacAvaney:2023:SIGIR, Upadhyay:2024b:arXiv, Faggioli:2023:ICTIR}.
While high correlation is often interpreted as evidence of LLM judge validity, this approach has significant limitations.

\end{itemize}

\issue{1}{Discriminative Ability and the Limits of Correlation}
Even though several studies report on strong leaderboard correlation between human and LLM judgments, \citet{Clarke:2024:arXiv} argue that Kendall's $\tau$ is ``less informative for assessing progress at the top of the leaderboard'' and demonstrate that LLM-based assessments fail to reliably identify the best-performing systems. Further, \citet{Alaofi:2024:SIGIRAP} show that correlation-based meta-evaluation hides interesting failure patterns.
A crucial, often overlooked, aspect is the disconnect between typical TREC evaluation setups and the needs of many practical IR scenarios.  TREC evaluations often involve dozens of systems with widely varying approaches and performance levels.  In contrast, practitioners often need to compare a small number of high-performing (state-of-the-art) systems or distinguish between subtle variations of a single system (e.g., in ablation studies).  It remains an open question whether LLM judges possess the necessary sensitivity to reliably detect small but meaningful performance differences in such scenarios.  Indeed, achieving high correlation is inherently easier with a larger and more diverse set of systems; simply including more systems with varying performance levels can artificially inflate correlation, even if the LLM judge struggles to differentiate between the top contenders. 
It thus remains an open question: \emph{Can LLM judges reliably distinguish between high-performing systems with small, but meaningful, performance differences?}

\subsubsection{Reliability and Reproducibility}

Beyond validity, a critical concern for LLM-based evaluation is the \emph{reliability} and \emph{reproducibility} of the judgments.  Even if an LLM demonstrates a reasonable level of agreement with human judgments on average, its utility as a judge is undermined if its assessments are highly sensitive to seemingly minor variations in setup or input.  Indeed, existing research demonstrates that LLM judgments can be significantly influenced by factors such as the choice of LLM~\citep{Faggioli:2023:ICTIR,Alaofi:2024:SIGIRAP,Chen:2024:SIGIRAP}, the specific wording and structure of the prompt~\citep{Alaofi:2024:SIGIRAP,Thomas:2024:SIGIR}, and even the order in which documents are judged~\citep{Chen:2024:SIGIRAP}. 
This variability raises concerns about the reliability of results obtained with a single LLM.

\issue{2}{The Impact of Model Choice} 
A recurring theme in the literature is that more powerful LLMs (typically larger models with more parameters and trained on larger datasets) tend to exhibit better performance and consistency as judges~\citep{Alaofi:2024:SIGIRAP}.  This raises a crucial, but largely unexplored, question: \emph{To what extent would the conclusions of a study change if a more (or less) powerful LLM were used as the judge?}  This sensitivity to model choice has not been systematically investigated, particularly in the context of comparing high-performing systems where subtle differences matter. 

\subsection{Vulnerability to Bias and Manipulation}
\label{sec:issues:bias}

Beyond the inherent quality of judgments, a separate set of concerns revolves around the potential for LLMs to be biased or manipulated, thereby impacting evaluation outcomes.

\subsubsection{Vulnerability to Manipulation}

A significant concern with the adoption of LLMs as judges is their potential vulnerability to adversarial manipulation.
Initial research suggests that LLM judges might be vulnerable to keyword stuffing and other SEO strategies~\citep{Alaofi:2024:SIGIRAP}. 
More broadly, knowledge of the (characteristics of the) LLM judge opens up ways to manipulate benchmarking results.  This could lead to situations where a system achieves much higher scores under automatic evaluation with the LLM judge than under manual assessment~\citep{Clarke:2024:arXiv}. 
This ``eval hacking'' undermines the purpose of evaluation, which is to accurately assess the \emph{true} utility of a system for users. 

\issue{3}{Understanding and Mitigating Vulnerabilities of LLM Judges}
While initial studies demonstrate the \emph{possibility} of manipulating LLM judges, the \emph{extent} of this vulnerability across different LLMs, attack strategies, and IR tasks remains largely unknown. \emph{What specific vulnerabilities do LLM judges exhibit, and how do these vulnerabilities vary across different models and evaluation settings?} Furthermore, \emph{How can we design evaluation protocols that are robust to manipulation, ensuring that LLM-based evaluation remains a reliable and trustworthy measure of system performance?} This is a crucial area for future research.

\subsubsection{Systematic Biases}

A core challenge in using LLMs for both ranking and evaluation lies in the fundamental similarity of the two tasks: both involve estimating the relevance of a document to a given query. Several studies note the potential for significant systemic biases when LLMs are employed in both roles~\citep{Rahmani:2024:arXiv,Faggioli:2023:ICTIR,Thomas:2024:SIGIR,MacAvaney:2023:SIGIR}. 
In their summary of the LLM4IR workshop, \citet{Rahmani:2024:arXiv} state ``if we were to use an LLM both as an assessor and as a ranker, we could expect such a model to be favoured over other evaluated models.''
\citet{Faggioli:2023:ICTIR} similarly caution that ``if the model is used to judge relevance both for annotation and for retrieval, its evaluation would be overinflated, possibly with perfect performance.''
If both ranking and automatic evaluation are predisposed towards certain types of results, it becomes difficult to identify \emph{truly} relevant results. This can lead to the suppression of diverse perspectives and the promotion of homogenous content. Novel ranking approaches that deviate from the LLM's inherent understanding of relevance might be unfairly penalized during the assessment phase. This phenomenon shares similarities with ``reward hacking'' observed in reinforcement learning, where agents exploit loopholes in the reward function to achieve high scores without genuinely solving the underlying task~\citep{Chen:2024:ICML}.
A particularly concerning form of this bias is \emph{circularity}, where retrieval models are trained on LLM-generated labels~\citep{Faggioli:2023:ICTIR,Rahmani:2024:arXiv,Clarke:2024:arXiv}. This creates a self-reinforcing loop, where the ranker learns to produce outputs that the LLM judge deems relevant, further amplifying any existing biases.

\issue{4}{Interrelated Systemic Biases in LLM-Based Evaluation}
While the potential for systemic biases in LLM-based IR evaluation is acknowledged, the specific interactions and magnitudes of these biases remain largely unquantified.  We identify three interrelated potential biases:

\begin{itemize}
    \item \emph{Bias Towards LLM-Based Rankers}: LLM judges might favor the output of systems that also employ LLMs for ranking. While intuitively plausible, this bias needs to be systematically investigated and quantified, independent of the content being retrieved. 
    \item \emph{Bias Towards LLM-Generated Text}: LLM judges might exhibit an inherent preference for text generated by LLMs, regardless of the ranking system that retrieved it. This could be due to factors like stylistic similarities, reduced noise, or other characteristics of LLM-generated text. Indeed, studies have observed that LLMs exhibit bias favoring texts generated by the same underlying model~\citep{Liu:2024:ACL,Panickssery:2024:NeurIPS}. However, there is a significant lack of studies systematically quantifying the extent to which LLM judges favor LLM-generated text in the specific context of IR evaluation.
    \item \emph{Combined Bias (LLM Ranker + LLM-Generated Text)}: The most complex scenario involves the potential interaction of the two biases above. 
    \citet{Dai:2024:KDD} show that neural retrievers prefer LLM-generated content, but their analysis relies on human judgments, not LLM judges.
    Does an LLM judge exhibit an even stronger preference for LLM-generated text when it is retrieved by an LLM-based ranker? This synergistic effect, if present, could significantly distort evaluation outcomes.
\end{itemize}
It thus remains a set of open questions: \emph{To what extent do LLM judges exhibit biases towards (1) LLM-based rankers, (2) LLM-generated text, and (3) the combination of the two?  How do these biases interact, and what is their combined impact on IR evaluation?}

\section{Experiments}
\label{sec:experiments}

To empirically demonstrate some of the challenges identified in Section~\ref{sec:issues}, we present a series of targeted experiments aimed at investigating the discriminative ability of LLM judgments (Issue \#1), the impact of model choice (Issue \#2), and systematic biases (Issue \#4).  Note that, our goal is to provide illustrative evidence of these issues, rather than a comprehensive or exhaustive analysis.

\subsection{Experiment Design}

We study the classic ad hoc retrieval task where a ranked list of documents are returned in response to a user query. We follow a standard retrieve-then-rerank paradigm, where an initial set of potentially relevant documents is identified by a fast and efficient first stage retriever, which are then subsequently reranked by a computationally more intensive but more accurate model.
Our focus lies specifically in this reranking stage, noting that LLMs may also be used for retrieval~\citep{Tay:2022:NeurIPS}.

We employ a set of rankers built upon progressively more capable LLMs. This allows us to observe how their performance changes as the underlying LLM technology advances and whether LLM judges indeed exhibit bias toward LLM-based rankers.
In a novel methodological approach, we also incorporate ``oracle'' rankings as reference points of comparison. These oracle rankings leverage ground truth human relevance labels to represent a hypothetical perfect ranking system as well as controlled degradations from this ideal. By intentionally degrading the perfect rankings, we create a spectrum of performance levels against which we can compare our LLM-based rankers as well as test the sensitivity of LLM judges.

For the judging side, we explore the sensitivity of evaluation by employing specific variations of LLM judges within a single model family---a relatively unexplored dimension in prior work.  By using these specific variations of LLM judges, we aim to assess the consistency and reliability of LLM-based evaluation and to understand how the choice of LLM judge might influence the predicted effectiveness of different rankers.  Crucially, we compare the judgments provided by these LLM judges against human assessments, which we consider as \emph{the} ground truth for relevance.

To further explore the implications of LLM integration across the information lifecycle, we also examine the impact of LLM-assisted content creation on retrieval and evaluation.  Specifically, we investigate how AI assistance in document authoring may influence relevance scores assigned by LLM-based rankers and judges.

\subsection{Experimental Setup}

We utilize the TREC Deep Learning (DL) 2019 and 2020 datasets~\citep{Craswell:2019:TREC,Craswell:2020:TREC}, chosen due to their extensive use in prior research in this area. 
Both use the MS MARCO v1 passage corpus, which contains 8.8 million passages. We adopt the convention of referring to passages as ``documents,'' even if the unit of retrieval are passages in our experiments.
The two datasets contain 43 and 54 queries, respectively, with human relevance annotations by TREC assessors. 

Following~\citep{Qin:2024:NAACL,Sun:2023:EMNLP}, all comparisons are based on the reranking of the top 100 passages retrieved by BM25~\citep{Lin:2021:SIGIR}. 
To ensure a fair comparison between human and LLM judges, we filter results that have not been judged by TREC assessors (instead of treating them as non-relevant).
For simplicity, we report only on NDCG@10, which is the official evaluation metric of the DL track.

\subsubsection{LLM Judges}
For automatic assessment, we use two model generations of a powerful commercial LLM, Gemini, in two sizes within each generation: v1 Nano, v1 Pro, v1.5 Flash, and v1.5 Pro.
We use the best applicable prompt\footnote{We use the prompt that considers multiple aspects (A), but not role (R) nor multiple judges (M); narrative (N) and description (D) are unavailable for TREC DL.} in \citep{Thomas:2024:SIGIR} based on the open source implementation UMBRELA~\citep{Upadhyay:2024:arXiv}, with judgments performed on a 4-point scale. 
We set top-p =1 and the temperature to 0.

\subsubsection{LLM Rankers}
We consider both supervised and unsupervised LLM-based rankers, in addition to a BM25 baseline:
\begin{itemize}
    \item \textbf{RankT5}~\citep{Zhuang:2023:SIGIR} is a reranker that uses T5~\cite{Raffel:2020:JMLR} and listwise ranking loss during supervised fine-tuning. It is considered a state-of-the-art supervised LLM-based ranker.
    \item \textbf{RG}~\citep{Liang:2023:TMLR} is a pointwise prompting method based on Relevance Generation, where the prompt asks ``Does the passage answer the query?'' and the logit of ``Yes'' is used as the ranking score. We test RG with FLAN-T5-XXL and FLAN-UL2. Note that RG requires internal logits of output tokens and thus cannot be used with black-box LLMs such as Gemini.
    \item \textbf{PRP}~\citep{Qin:2024:NAACL} is a pairwise prompting approach that is effective and robust across LLMs with different sizes. Given a query and two passages, the prompt asks ``Which of the two passages is more relevant to the query?'' The winning rate is used as the ranking score for each passage.
\end{itemize}

\begin{figure}[t]
    \centering
    \includegraphics[width=0.38\textwidth]{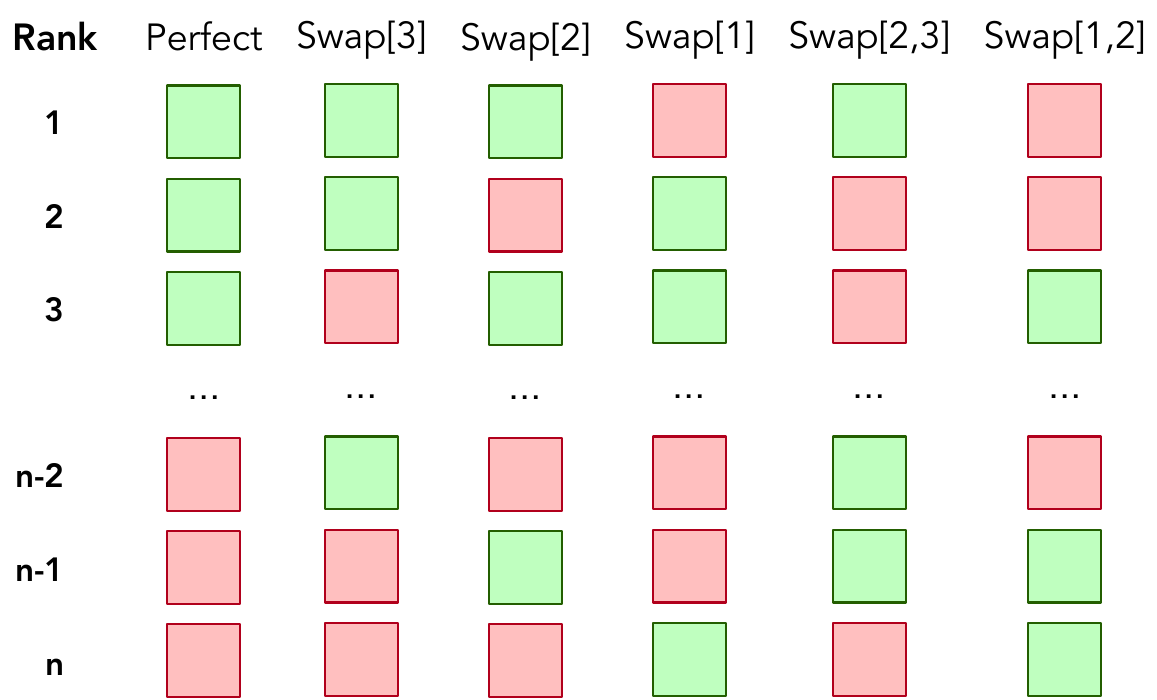}
    \vspace*{-0.25\baselineskip}
    \caption{Illustration of oracle rankers, ordered by their expected performance, assuming that the top three results are highly relevant and the bottom three are non-relevant.}
    \label{fig:oracles}
    \vspace*{-0.25\baselineskip}
\end{figure}

\begin{table*}[t]
    \centering
    \caption{Results (NDCG@10) on the TREC DL 2019 and 2020 collections using both human and LLM judges. The best LLM and Oracle reranking approaches per judge are boldfaced.}
    \label{tab:results}
    \vspace*{-0.5\baselineskip}
    \small
    \begin{tabular}{llccccclccccc}
    \toprule
    \multirow{3}{*}{\textbf{Method}} & \multirow{3}{*}{\textbf{LLM}} & 
        \multicolumn{5}{c}{\textbf{TREC DL19}} & & 
        \multicolumn{5}{c}{\textbf{TREC DL20}} \\
    \cline{3-7} \cline{9-13}
    & & 
        \textbf{Human} & \multicolumn{4}{c}{\textbf{LLM judges}} & &
        \textbf{Human} & \multicolumn{4}{c}{\textbf{LLM judges}} \\
        \cline{4-7} \cline{10-13}
    & & 
        \textbf{judges} & \textbf{\scriptsize{v1 Nano}} & \textbf{\scriptsize{v1 Pro}} & \textbf{\scriptsize{v1.5 Flash}} & \textbf{\scriptsize{v1.5 Pro}} & &
        \textbf{judges} & \textbf{\scriptsize{v1 Nano}} & \textbf{\scriptsize{v1 Pro}} & \textbf{\scriptsize{v1.5 Flash}} & \textbf{\scriptsize{v1.5 Pro}} \\
    \midrule
    \multicolumn{2}{l}{\emph{Initial retrieval}} \\
    \midrule
    BM25 & - & 0.506 & 0.607 & 0.772 & 0.689 & 0.712 & 
        & 0.483 & 0.616 & 0.786 & 0.689 & 0.719 \\
    \midrule
    \multicolumn{2}{l}{\emph{LLM reranking}} \\
    \midrule
    RankT5 & T5 (3B) & 0.731 & \textbf{0.633} & 0.907 & 0.911 & 0.916 & 
        & 0.696 & 0.621 & 0.924 & 0.888 & 0.899 \\
RG & FLAN-T5-XXL (11B) & 0.673 & 0.606 & 0.895 & 0.874 & 0.881 & 
    & 0.639 & 0.619 & 0.920 & 0.877 & 0.878 \\
 & FLAN-UL2 (20B) & 0.689 & 0.595 & 0.896 & 0.884 & 0.887 & 
    & 0.667 & 0.611 & 0.922 & 0.880 & 0.885 \\
PRP & FLAN-T5-XL (3B) & 0.716 & 0.610 & 0.924 & 0.921 & 0.909 & 
    & 0.691 & 0.618 & 0.924 & 0.898 & 0.901 \\
 & FLAN-T5-XXL (11B) & 0.712 & 0.620 & 0.918 & 0.922 & 0.926 & 
    & 0.712 & 0.615 & 0.938 & 0.905 & 0.912 \\
 & FLAN-UL2 (20B) & 0.734 & 0.614 & 0.923 & 0.914 & 0.928 & 
    & \textbf{0.718} & \textbf{0.622} & 0.932 & 0.909 & 0.917 \\
 & Gemini v1.5 Flash & \textbf{0.747} & 0.623 & \textbf{0.937} & \textbf{0.961} & \textbf{0.947} & 
    & 0.699 & 0.619 & \textbf{0.952} & \textbf{0.937} & \textbf{0.933} \\
    \midrule 
    \multicolumn{2}{l}{\emph{Oracle reranking}} \\
    \midrule 
    Perfect & - & \textbf{0.892} & 0.582 & \textbf{0.896} & \textbf{0.876} & \textbf{0.864} & 
        & \textbf{0.871} & 0.617 & \textbf{0.872} & \textbf{0.828} & \textbf{0.824} \\
Swap[3] & - & 0.824 & 0.589 & 0.868 & 0.835 & 0.827 & 
    & 0.795 & 0.611 & 0.842 & 0.795 & 0.796 \\
Swap[2] & - & 0.814 & 0.589 & 0.859 & 0.825 & 0.814 & 
    & 0.790 & 0.611 & 0.853 & 0.797 & 0.797 \\
Swap[1] & - & 0.803 & 0.578 & 0.870 & 0.836 & 0.836 & 
    & 0.764 & \textbf{0.621} & 0.832 & 0.778 & 0.776 \\
Swap[2,3] & - & 0.739 & \textbf{0.596} & 0.829 & 0.779 & 0.771 & 
    & 0.706 & 0.602 & 0.821 & 0.760 & 0.765 \\
Swap[1,2] & - & 0.713 & 0.585 & 0.831 & 0.782 & 0.783 & 
    & 0.672 & 0.615 & 0.810 & 0.743 & 0.746 \\
\bottomrule
    \end{tabular}
\end{table*}

\begin{figure*}[t]
    \centering
    \begin{tabular}{@{}cc@{}}
        \includegraphics[height=4.1cm]{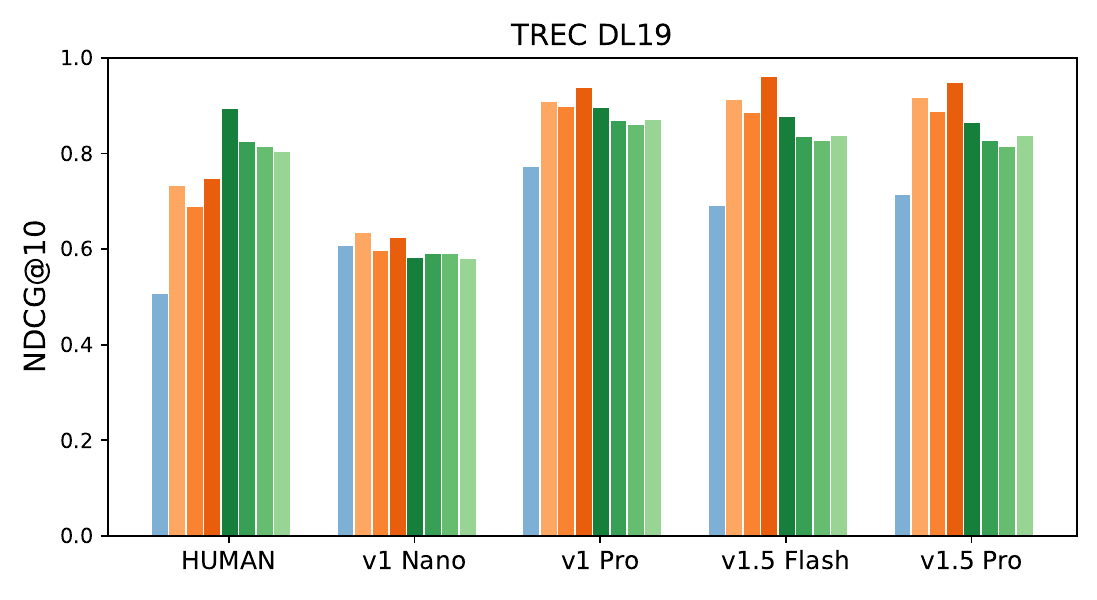}
         &  
        \includegraphics[height=4.1cm]{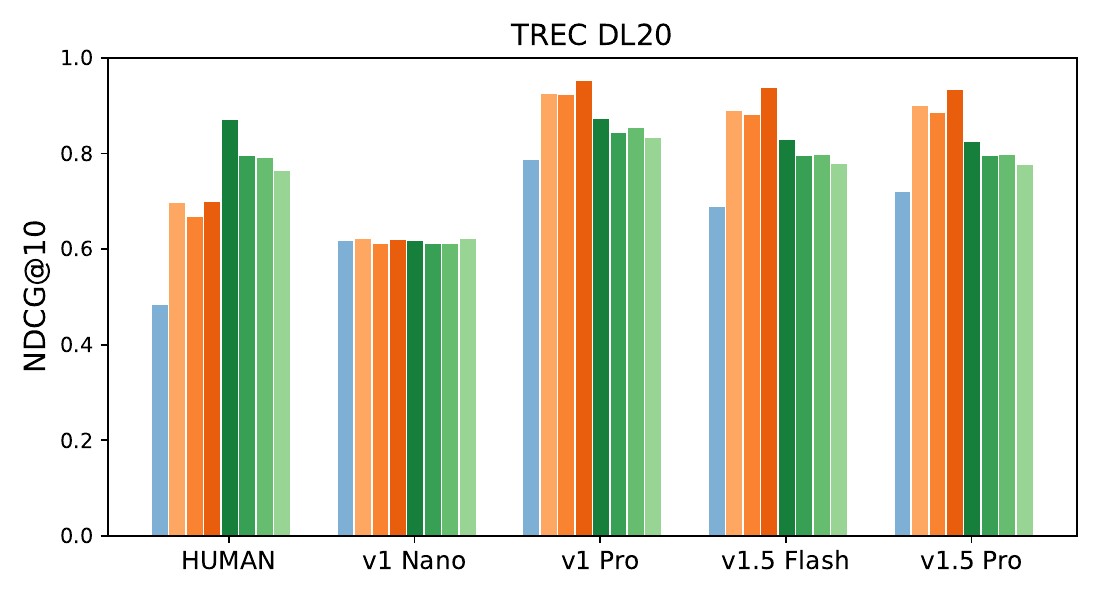}
        \\         
    \end{tabular}
    \vspace*{-1.25\baselineskip}
    \caption{Visualization of the performance of selected rankers from Table~\ref{tab:results}.}
    \label{fig:results}
\end{figure*}

\subsubsection{Oracle Rankers}
We generate oracle rankings using the ground truth TREC relevance assessments.
To ensure a fair comparison with LLM rankers, we consider the same initial set of BM25-retrieved documents for reranking. 
Specifically, we consider the following oracle rankers, which are visualized in Fig.~\ref{fig:oracles}:
\begin{itemize}
    \item{\textbf{Perfect}}: Reranks results according to the ground truth relevance labels. While not perfect overall, this represents the ideal ranking within the initially retrieved set.
    \item{\textbf{Swap[i]}}: Introduces controlled errors by swapping the top-$i$ ranked result with the bottom-$i$ result. Decreasing $i$ (from 3 to 2 to 1) increases the deviation from the perfect ranking.
    \item{\textbf{Swap[i,i+1]}}: Swaps the $i$th and ($i+1$)th highest-ranked results with the $i$th and ($i+1$)th lowest-ranked results. This represents further degradation from the Swap[i] methods.
\end{itemize}

\subsubsection{Measuring Alignment}
Following prior work (cf. Section~\ref{sec:issues:validity}) we measure agreement with TREC judges (on all human-judged query-document pairs) in terms of Cohen's $\kappa$ using both graded and binary relevance labels. Following \citet{Faggioli:2023:ICTIR}, we create binary relevance labels by merging levels 0 and 1 (non-relevant) and levels 2 and 3 (relevant).
Additionally, we report on relative system ordering in terms of Kendall's $\tau$.

\begin{table*}[t]
    \centering
    \caption{Agreement between LLM and human judges (1) on individual relevance judgments (Cohen's $\kappa$) using both graded and binary labels and (2) on relative ordering of systems (Kendall's $\tau$) considering all systems in Table~\ref{tab:results} and Oracle rankers only. 
    }
    \label{tab:correlation}
    \vspace*{-0.5\baselineskip}
    \small
    \begin{tabular}{lccccccccccc}
    \toprule
    \multirow{3}{*}{\textbf{LLM judge}} & \multicolumn{5}{c}{\textbf{Cohen's} $\kappa$} && \multicolumn{5}{c}{\textbf{Kendall's} $\tau$} \\
    \cline{2-6} \cline{8-12}
    & \multicolumn{2}{c}{Graded} && \multicolumn{2}{c}{Binary} && \multicolumn{2}{c}{All systems} && \multicolumn{2}{c}{Oracles-only} \\
    \cline{2-3} \cline{5-6} \cline{8-9} \cline{11-12}
    & DL19 & DL20 && DL19 & DL20 && DL19 & DL20 && DL19 & DL20 \\
    \midrule
    v1 Nano & -0.002 & -0.011 && 0.007 & -0.003 && -0.253 & 0.011 && -0.067 & 0.067 \\
    v1 Pro & 0.139 & 0.144 && 0.337 & 0.273 && \textbf{0.077} & 0.121 && \textbf{0.600} & \textbf{0.867} \\
    v1.5 Flash & \textbf{0.268} & \textbf{0.230} && 0.461 & \textbf{0.370} && 0.033 & \textbf{0.143} && \textbf{0.600} & \textbf{0.867 }\\
    v1.5 Pro & 0.204 & 0.192 && \textbf{0.462} & 0.359 && \textbf{0.077} & \textbf{0.143} && \textbf{0.600} & \textbf{0.867} \\    
    \bottomrule
    \end{tabular}
\end{table*}

\subsection{Results}

Table~\ref{tab:results} presents the results of the various reranking methods evaluated using both human and LLM judges. Selected methods are shown in Fig.~\ref{fig:results} for easier visual inspection. Additionally, Table~\ref{tab:correlation} reports on agreement between human and LLM judges. 

\boldheading{Choice of LLM}
\emph{How well do LLM-based judgments align with human assessments when using different variations of LLM judges from the same model family?}
Looking at the evaluation scores of various rankers, we observe generally good agreement among the three largest models. In terms of agreement with human judges on individual relevance judgments (see Cohen's $\kappa$ Table~\ref{tab:correlation}) the results are comparable to those reported in prior work for these datasets~\citep{Upadhyay:2024:arXiv}, with the newer v1.5 models performing clearly better than the v1 models. Interestingly, within this newer model generation, a larger model is not necessarily more capable, at least not according to this measure; v1.5 Flash shows much better agreement with humans when a graded relevance scale is used than the v1.5 Pro.
On the other hand, the smallest LLM (v1 Nano) is unable to provide useful judgments, as evidenced by the Cohen's $\kappa$ values being close to 0.  While this model may be capable in other tasks~\citep{Gemini:2023:arXiv}, our results clearly indicate its unsuitability for judging relevance in this specific context. Therefore, we exclude the v1 Nano judge from subsequent analyses and discussions referring to ``LLM judges.''

Another way to validate LLM judges is by measuring how well they agree on the relative ordering of systems with human judges; see Kendall's $\tau$ Table~\ref{tab:correlation}. In this regard, the newest and largest model (v1.5 Pro) is the most capable overall, but there is in fact little difference in performance among the three largest models. Thus, while newer model generations clearly perform better (v1 vs. v1.5), larger models with the same generation do not necessarily make more capable judges (v1.5 Flash vs. v1.5 Pro). We also note that differentiating between the entire pool of systems (``All systems'') proves to be especially challenging; we will elaborate on this next.

\boldheading{Discriminative Ability}
\emph{Can LLM judges reliably distinguish between high-performing systems with small, but meaningful, performance differences?}
The Oracle rankers, with their controlled performance degradations, enable us to assess the discriminative power of LLMs in a setting free from potential biases introduced by LLM-based rankers. 
While the absolute score differences between some pairs of Oracle rankings may be small, all pairwise differences are statistically significant according to human judgments (paired t-test, $p<0.05$). 
Therefore, a failure to observe a statistically significant difference, or, more critically, a reversal of the correct ordering, indicates that the LLM judge is not sufficiently sensitive.
Table~\ref{tab:correlation} (Oracle-only setting) reveals that accurately ordering the Oracle rankings is challenging for LLM judges, particularly on the DL19 dataset. This suggests limitations in their ability to discern subtle, yet statistically significant, performance differences.
This limited discriminative ability is not confined to the Oracle setting; it also manifests when evaluating actual retrieval systems. For instance, the v1.5 Pro model, which performed best among the LLMs on the Oracle rankings, fails to identify statistically significant differences between certain pairs of systems (e.g., RankT5 vs. RG-FLAN-T5-XXL on DL19, $p<0.001$ according to human evaluation). Conversely, it can also identify differences as statistically significant (e.g., PRP-FLAN-UL2 vs. PRP-Gemini-v1.5-Flash, $p<0.05$ for both years) when human judgments show no significant difference.

Furthermore, the substantial difference in correlation between the ``All systems'' and ``Oracles-only'' results in Table~\ref{tab:correlation} provides direct evidence of the concerns raised in Issue \#1 (Section \ref{sec:issues:validity}), namely, how easily correlation-based metrics can be manipulated by the choice of systems included in the evaluation.

\boldheading{Bias Towards LLM-based Rankers}
\emph{Do LLM judges exhibit biases towards LLM-based rankers?}
The results presented in Fig.~\ref{fig:results} demonstrate a clear and substantial bias in favor of LLM-based rankers when evaluated by LLM judges.
While prior work has hinted at the \emph{potential} for such a bias, this study provides direct empirical evidence of its existence and magnitude.
Human judgments consistently place the selected Oracle rankers shown in Fig.~\ref{fig:results} above all LLM-based rankers. LLM judges, however, completely invert this order, ranking \emph{all} LLM-based rankers as superior to these Oracle runs. This is not a subtle effect; the magnitude of the bias is sufficient to completely reverse the relative ranking of these two fundamentally different types of systems.  The fact that the true performance of non-LLM-based systems is severly underestimated, highlighting a critical limitation of relying solely on LLM judges for evaluation, particularly when assessing fundamentally new or unconventional approaches.

\begin{figure}[t]
    \centering
    \includegraphics[width=0.45\textwidth]{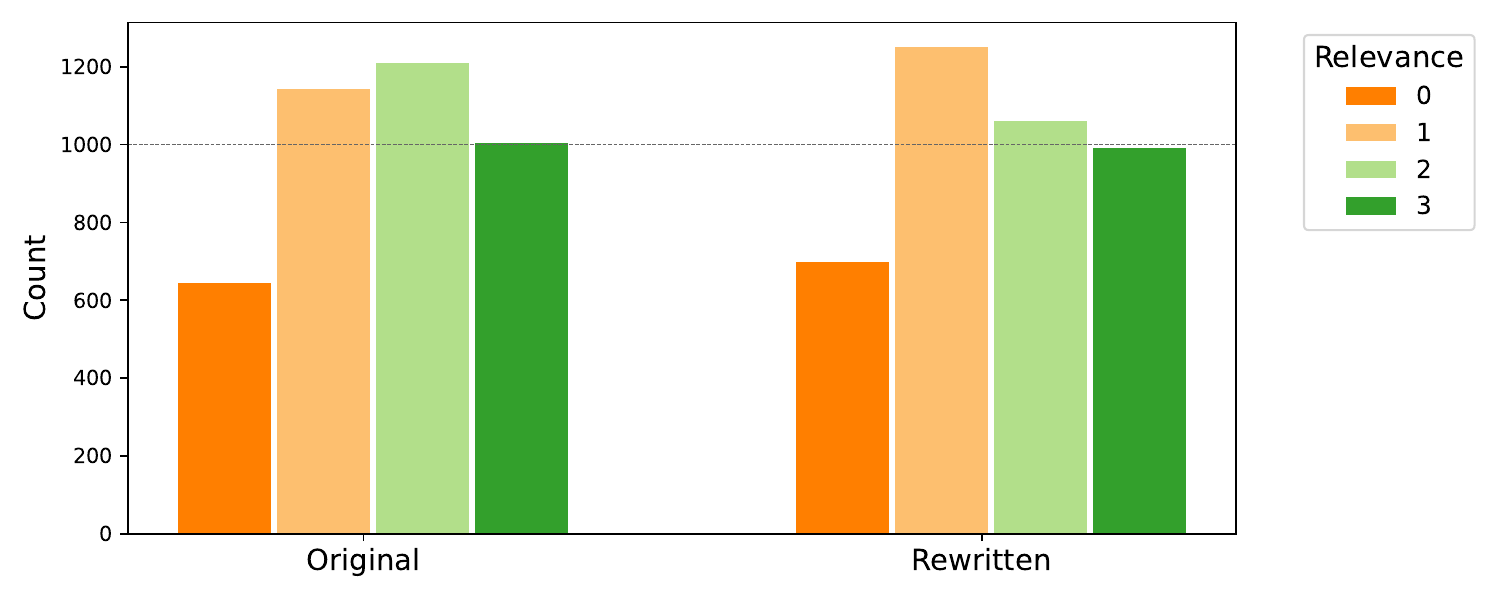}
    \vspace*{-1\baselineskip}
    \caption{Relevance levels estimated by an LLM judge (Gemini v1.5 Pro) for Original vs. AI-assisted content (Rewrites, using Gemini v1.5 Flash).  According to human assessors, the labels should be uniformly distributed across the four relevance classes, as indicated by the dashed horizontal line.}
    \label{fig:rewrites}
    \vspace*{-0.5\baselineskip}
\end{figure}

\boldheading{Bias Towards LLM-generated Text}
\emph{Do LLM judges exhibit biases towards LLM-generated text (independent of the ranking mechanism used to retrieve that text)?}
We investigate this by comparing LLM judge assessments of original human-written documents and their AI-assisted counterparts. Using the MS MARCO dataset, which predates the widespread adoption of modern AI writing tools, we can reasonably assume that the original documents represent content created without significant AI assistance. 
To avoid bias potentially introduced by initial retrieval, we take a balanced sample: for each year (DL19 and DL20) we randomly sample 500 query-document pairs for each of the four relevance levels, resulting in a total of 4000 query-document pairs. 
We refer to this set as Original.
We then employ our second most capable LLM (Gemini v1.5 Flash) to create an AI-rewritten version of each document in the Original set, following the methodology of \citet{Dai:2024:KDD}.\footnote{They use the straightforward prompt ``Please rewrite the following text: {{human-written text}}'' in a zero-shot setting.} 
This rewritten set is referred to as Rewritten.
We shall assume that this rewriting process does not substantially alter the relevance of the documents to their corresponding queries, as verified by human assessors in~\citep{Dai:2024:KDD}.
Figure~\ref{fig:rewrites} presents the results using our most capable LLM (Gemini v1.5 Pro) as the judge.
We can observe on the Original data that the LLM judge is lenient in its assessment of relevance, and specifically in labeling non-relevant documents as partially relevant. However, the judge does not appear to systematically inflate scores for the highest relevance level.  Crucially, when comparing these results to the judgments on the Rewritten (LLM-generated) text, we do not observe a distributional shift towards higher relevance levels. In fact, the Rewritten documents show a slight increase in lower relevance labels. While these findings are specific to this particular combination of LLM rewriter and judge, they provide evidence against a general bias towards LLM-generated content, even when both models are from the same family.

It is important to note that the preceding analysis examines the \emph{distributional} impact of LLM-generated text on relevance judgments. To further investigate potential biases in a ranking context, we conduct a second experiment. We take the rankings produced by the perfect Oracle method and re-evaluate them using our LLM judge (Gemini v1.5 Pro). However, instead of using the Original document content, we substitute the Rewritten versions.  If the LLM judge exhibited a strong preference for LLM-generated text, we would expect to see a significant increase in the scores assigned to these rankings. However, according to our results, that is not the case. We find that the performance of the Perfect Oracle method, as assessed by the LLM judge, does not change significantly when using the Rewritten text instead of the Original text: we get an NDCG@10 of 0.868 vs. 0.883 on DL19 and 0.825 vs. 0.818 on DL20 for Rewritten vs. Original; none of these differences is statistically significant.
This further reinforces the conclusion that, at least in this experimental setup, the LLM judge does not exhibit a strong bias towards LLM-generated content.

\begin{figure}[t]
    \centering
    \includegraphics[width=0.45\textwidth]{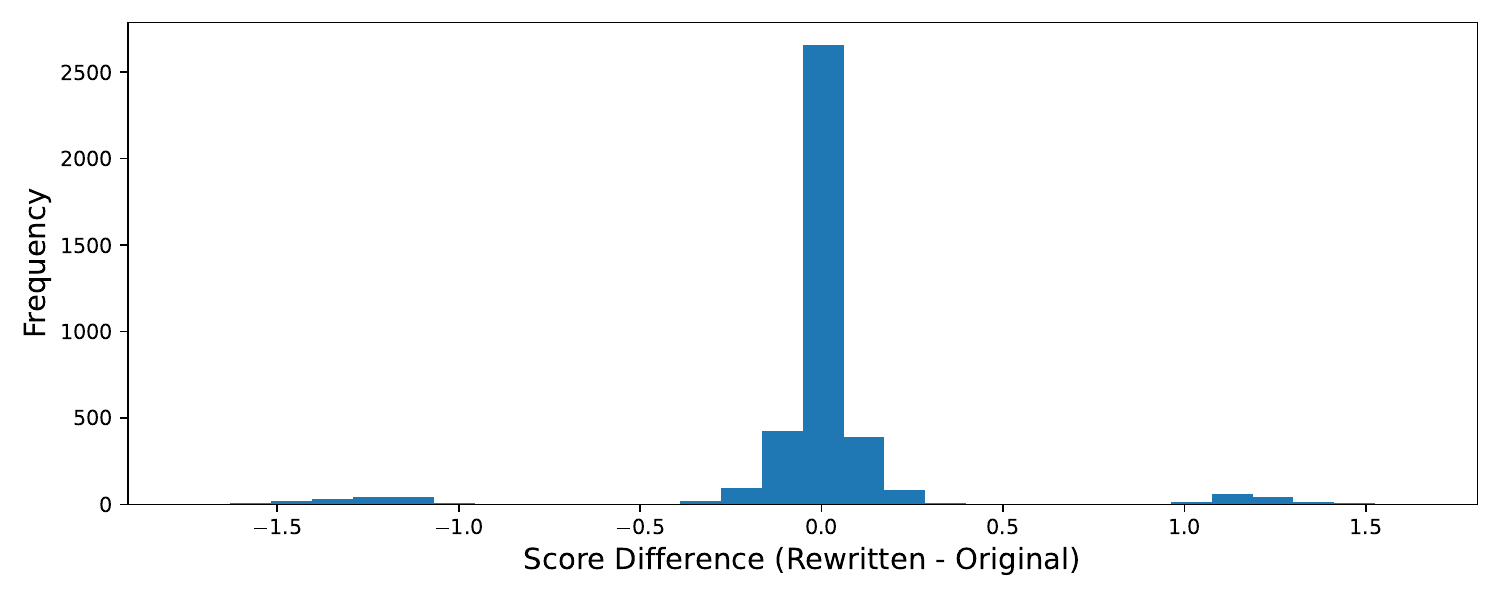}
    \vspace*{-1\baselineskip}
    \caption{Distribution of score differences of a RankT5 ranker on LLM-rewritten vs. original text on a sample of 4000 query-document pairs.}
    \label{fig:rewrites_rankt5}
    \vspace*{-0.5\baselineskip}
\end{figure}

\boldheading{Combined Bias (LLM Ranker + LLM-Generated Text)}
\emph{Do LLM judges exhibit biases towards LLM-generated text when using LLM-based rankers?}
To address this question, we conduct an experiment combining an LLM-based ranker with LLM-generated text and an LLM judge. We utilize the same balanced sample of 4,000 query-document pairs (500 per relevance level for each of DL19 and DL20) used in the previous experiment, comprising both the Original and Rewritten sets.
For ranking, we employ a pointwise approach using RankT5 with a Flan-T5-XXL model. We compare the scenarios where the RankT5 model scores (1) the Original query-document pairs and (2) the Rewritten query-document pairs. Both scorings are then evaluated using the same LLM judge (Gemini v1.5 Pro). 
We observe minimal differences in the LLM-assigned evaluation scores between the Original and Rewritten scenarios.  Closer inspection of the RankT5 scores reveals that the rewriting process had a negligible impact on retrieval scores for the vast majority of query-document pairs. The few observed changes were symmetrically distributed, with increases and decreases in scores mirroring each other; see Fig.~\ref{fig:rewrites_rankt5}.
This aligns with the previous experiment's findings, suggesting that neither the LLM judge nor the LLM ranker (in this specific configuration) exhibits a strong preference for the LLM-rewritten content. Consequently, the LLM judge produces very similar evaluation results in both cases.
While these results do not demonstrate a combined bias in this specific experimental setup, the potential for synergistic effects between LLM rankers, LLM-generated text, and LLM judges remains an open question requiring further, more comprehensive investigation.

\section{The Role and Challenges of LLM Judges in IR}
\label{sec:challenges}

The feasibility of LLMs as automatic relevance assessors has been established and they have been rapidly adopted both in academia and in industry. 
The question, therefore, is not whether they \emph{can} be used as judges, but rather \emph{how} they should be used in a principled and effective manner. 
This requires a careful consideration of both the intended purpose of LLM judges and their inherent limitations.

\boldheading{Clarifying the Purpose}
Most studies, albeit often implicitly, employ LLMs as judges with the aim of \emph{replacing} human assessors. However, as we discuss below, fundamental limitations preclude this possibility.  We argue (in line with~\citep{MacAvaney:2023:SIGIR,Faggioli:2023:ICTIR}) that a more appropriate and productive goal should be to enable more effective use of limited human assessor time and resources. This shift in perspective---from replacement to reducing human effort---is crucial for guiding the development and deployment of LLM judges.

\boldheading{Acknowledging Fundamental Limitations} 
It is essential to recognize that both ranking and relevance assessment address the same problem: predicting the relevance of a document to a given query. This inherent overlap introduces fundamental limitations when using LLMs for both tasks.
\citet{Clarke:2024:arXiv} argue that ``A true gold standard must originate from human assessments, as only humans can determine the relevance of information in a way that reflects real-world utility.'' 
We must also recognize that relevance itself carries an intrinsic uncertainty; it depends on the entire cognitive state of the person, which changes as they use the system~\citep{Saracevic:1996:COLIS}.
Because of these inherent limitations, ``LLM assessments may themselves represent a strong ranking method, rather than a valid evaluation metric''~\citep{Clarke:2024:arXiv}. Recent work, such as \citep{Oosterhuis:2024:KDD}, has begun to provide uncertainty measures with LLM relevance predictions.

\boldheading{What, then, is the Role of LLM Judges?}
Given the limitations outlined above, and recognizing that information access systems are ultimately built to serve \emph{human} needs and provide utility to \emph{users}, the highest-fidelity evaluation of these systems can only be achieved through online evaluation with real users. Offline evaluation, while valuable, remains an abstraction of the real-world task because it removes the user from the evaluation process. There is a genuine risk that findings from offline experiments, particularly those relying solely on LLM judges, may not translate to operational settings. It is crucial to recognize that the signal provided by LLM judges is inherently a noisy and potentially biased one, and therefore cannot be fully trusted as a direct proxy for utility. Nevertheless, this noisy signal can be a useful indicator, helping to identify which methods or system variants are promising enough to warrant the more resource-intensive process of human evaluation.

\subsection{Guidelines for Employing LLMs as Judges}

We outline several key considerations for employing LLMs in evaluation, aiming to foster a community-wide set of best practices and ensure methodological soundness. These are not exhaustive, but represent an important starting point.

\begin{itemize}
    \item \emph{Consistent Evaluation Across Systems.} All systems being compared within a single evaluation should be assessed using the same LLM judge configuration (model, prompt, settings). This ensures a fair and unbiased comparison, avoiding situations where some systems are evaluated with a more lenient or biased judge than others. Specifically, LLM judges should not be used selectively to fill ``relevance holes'' in existing human judgments~\citep{Abbasiantaeb:2024:EMTCIR}.
    \item \emph{Transparency and Reproducibility.} To enable reproducibility and facilitate comparisons across studies, researchers should clearly report the specific LLM used (model version), the exact prompt(s) employed, and any relevant settings or parameters.
    \item \emph{Employing Multiple LLMs as Judges.} Using a combination of different LLM judges can help mitigate biases stemming from LLMs favoring responses from their own model family and improve robustness (see, e.g.,~\citep{Jacovi:2025:arXiv}). Reporting the distribution of scores across different judges, as suggested by \citet{Rahmani:2024:arXiv}, can further enhance robustness.
    \item \emph{Alignment with Human Preferences.} Ensuring alignment between human and LLM raters is a substantial effort that needs to be continuously monitored and refined~\citep{Thomas:2024:SIGIR}. Ideally, results reported on LLM judges should also include human validation of the results on a representative sample. Researchers should also exercise care when making research claims based on results from LLM judges.
\end{itemize}

\subsection{Open Questions and Future Directions}

The adoption of LLMs as judges in IR presents several open questions and necessitates further research to address the limitations and biases identified in previous sections.

\begin{itemize}
    \item \emph{Assessing and Improving LLM Judge Quality.} Our findings highlight the critical importance of LLM judge quality, revealing shortcomings in their discriminative ability and biases toward LLM-powered rankers. Developing robust methods for assessing and improving the quality of LLM judges is a crucial research direction for the IR community, potentially drawing motivations from horizontal autorater efforts~\citep{Vu:2024:EMNLP}. 
    \item \emph{Robustness Against Adversarial Attacks.}  LLM judges, similar to LLM-based rankers, are susceptible to adversarial attacks, including keyword stuffing and content injection~\citep{Alaofi:2024:SIGIRAP,Tamber:2025:arXiv}. Understanding these vulnerabilities and developing effective mechanism to enhance the robustness of LLM judges against them are critical areas for ensuring the practical applicability of LLM judges in real-world scenarios.
    \item \emph{Human-in-the-Loop LLM Judges.} The potential for using LLMs to augment human assessors, e.g., for quality-control, has been suggested~\citep{Faggioli:2023:ICTIR,Soboroff:2025:IRR}. While \citet{Upadhyay:2024b:arXiv} found that human-in-the-loop processes did not bring obvious tangible benefits, their study represents a preliminary investigation. Further research is needed to explore the potential of this approach more comprehensively.
    \item \emph{From Passages to Longer Documents.} Most of the existing work focuses on paragraphs as the unit of retrieval, using either the TREC DL~\citep{MacAvaney:2023:SIGIR,Faggioli:2023:ICTIR,Alaofi:2024:SIGIRAP,Chen:2024:SIGIRAP} or RAG~\citep{Upadhyay:2024b:arXiv} benchmarks. There is much less work on ad hoc retrieval, with exceptions including TREC-8~\citep{Faggioli:2023:ICTIR} and Robust~\citep{Thomas:2024:SIGIR}. It is known that LLMs handle long context differently~\citep{Liu:2024:TACL} and its implication in judging long documents need further investigation. 
    \item \emph{Alternative Judging Approaches.} All existing studies apply LLMs in a pointwise manner, but a pairwise or listwise setup would also be possible. LLM judge research can borrow ideas from LLM ranking research where pairwise and listwise approaches are extensively explored.
    \item \emph{Domain-specific Solutions.} While most existing research focuses on general-purpose search, specific domains might require purpose-built solutions. Recent work, for example, has explored the use of LLM judges for e-commerce search~\citep{Mehrdad:2024:arXiv,Sachdev:2025:arXiv}. Extending this concept of domain specialization, the applicability and value of LLM judges in specialized domains requiring expert knowledge (e.g., medical or legal search) remain less clear. In such domains, LLM judges might offer potential cost savings and more in-depth domain-specific knowledge compared to non-expert human assessors, but they also introduce new challenges, including the need for high accuracy, the potential for serious consequences from errors, and the complexities of expert judgment. Future research should explore the use of LLM judges in these contexts, carefully considering the trade-offs between cost, efficiency, and the risks of inaccurate assessments.
    \item \emph{Smaller, Purpose-Built Models.} The use of LLMs for judging at large scale raises concerns about computational cost and latency. A promising research direction is to explore the development of smaller models, designed specifically for the judging task. These purpose-built models could potentially offer significant advantages in terms of efficiency and speed, while maintaining performance comparable to massive LLMs in terms of accuracy and reliability.
    \item \emph{Internationalization.} Most related research focuses on English-language corpora. The issues discussed in this paper may be amplified in other languages, especially low resource ones, due to limitations in LLMs' multilingual capabilities. Further research is needed to evaluate the performance of LLM judges across a wider range of languages.
    \item \emph{Training Models on LLM-generated Labels.}  Training retrieval models on data labeled by LLMs introduces a significant risk of circularity and bias amplification. If done recursively, this might lead to model collapse~\citep{Shumailov:2024:Nature}.  \citet{Thomas:2024:SIGIR} acknowledge that parts of the Bing search engine are retrained using LLM-generated labels. Understanding the long-term effects of such training is an important research direction.
\end{itemize}

\section{Conclusion}
\label{sec:concl}

This paper has investigated the emerging and critical challenge of understanding the effect LLM-based rankers and AI-powered content creation may have on LLM-based judges' ability to accurately assess relevance. 
Through a synthesis of existing literature, we identified key concerns regarding the quality, validity, reliability, and potential biases of LLM judgments.  Our experiments provided empirical evidence demonstrating how interactions between the various roles LLMs play can lead to inaccurate or biased assessments of retrieval effectiveness, particularly in scenarios involving LLM-based rankers.
Finally, we presented guidelines for the use of LLMs as judges in IR and outlined a research agenda to address crucial open questions in this rapidly evolving field.

\balance
\bibliographystyle{ACM-Reference-Format}
\bibliography{sigir2025-llms.bib}

\end{document}